\documentclass[aps,pre,twocolumn,showpacs,amsmath,amssymb]{revtex4-1}
\usepackage{graphicx}
\usepackage{amsmath}

\usepackage{graphicx}

\begin{document}

\title{Control of transport in higher dimensional systems via dynamical 
decoupling of degrees of freedom with quasiperiodic driving fields}

\author{David Cubero$^{1}$ and Ferruccio Renzoni$^{2}$ }

\affiliation{$^{1}$Departamento de F\'{\i}sica Aplicada I, EUP, Universidad 
de Sevilla, Calle Virgen de \'Africa 7, 41011 Sevilla, Spain}
\affiliation{$^{2}$Department of Physics and Astronomy, University College
London, Gower Street, London WC1E 6BT, United Kingdom}

\date{\today}

\begin{abstract}
We consider the problem of the control of transport in higher
dimensional periodic structures by applied ac fields. In a generic 
crystal, transverse degrees of freedom are coupled, and this makes
the control of motion difficult to implement. We show, both with 
simulations and with an analytical functional expansion on the 
driving amplitudes, that the use of quasiperiodic driving significantly 
suppresses the coupling between transverse degrees of freedom.
This allows a precise control of the transport, and does not require 
a detailed knowledge of the crystal geometry. 
\end{abstract}


\maketitle

\section{Introduction}

 Periodic and quasiperiodic structures, both in time and in space,
exhibit completely different properties. For the case of spatial
quasiperiodicity, it is well established that quasiperiodic crystals exhibit 
properties which are very different from their periodic counterpart. In 
particular, transport properties, which are the main focus of this work, are 
significantly modified in the transition from a periodic structure to a 
quasi-periodic one. The transition from a ballistic regime in a periodic 
crystal to a regime of anomalous diffusion in a perfect quasicrystal well 
highlights the profound difference between the two structures. Mathematically,
a quasicrystal can be treated as a periodic structure embedded in an hyperspace
of higher dimension. That is, the effective dimensionality of the system is 
changed in the transition from periodicity to quasiperiodicity. This is the 
feature that, in the time-domain, will be central to our analysis.

In this work we consider the problem of the control of transport in
higher-dimensional crystals via ac driving fields \cite{reimann02,hanmar09}. 
In a generic crystal transverse degrees of freedom are coupled, 
and this makes the control of motion difficult to implement. Inspired by the 
above unique feature of quasiperiodic structures, we examine the case of a 
periodic spatial lattice and a quasiperiodic driving. We demonstrate, 
both with simulations and with a quite general functional expansion on the 
driving amplitudes, that the 
use of quasiperiodic driving leads to a {\it dynamical decoupling} of degrees 
of freedom, whereby the coupling between transverse degrees of freedom is 
significantly suppressed. This allows a precise control of the transport, 
independently of the lattice structure. 

\section{Model and definitions}
In the simulations, we  choose as an example the  dynamics of  a classical  
particle described  by the Langevin equation
\begin{equation}\label{eq:motion}
m\ddot{{\bf   r}}=   -\alpha\dot{{\bf   r}}-\nabla   U({\bf   r})+{\bf
F}(t)+{\bf \xi}(t),
\end{equation}
where ${\bf r}=(x,y)$ is the coordinate vector of the particle, $m$ is
its mass, $\alpha$ the friction coefficient, ${\bf \xi}=(\xi_x,\xi_y)$
a fluctuating force modeled by two independent Gaussian white noises,
$\langle
\xi_i(t)\xi_j(t^\prime)\rangle=2D\delta(t-t^\prime)\delta_{ij}$
($i,j=x,y$),  ${\bf F}(t)$  an applied time-dependent  driving to  be
specified later on, and  $U({\bf r})$ a two-dimensional space-periodic
potential that is also spatially symmetric in both directions $x$ and $y$. 
We have considered first the potential
\begin{equation}\label{eq:rect_pot}
U({\bf r})=U_0\cos(k x)[1+\cos(2ky)],
\end{equation}
which defines a {\em rectangular} lattice. 
Throughout   the   paper,   reduced   units  are   assumed  so  that
$m=k=U_0=1$. In  these units, the  friction coefficient and  the noise
strength were fixed to $\alpha=0.1$ and $D=0.5$. 

 This system model contains noise, dissipation, and finite inertia, which are important ingredients in the modelling of the experiments using 2D optical lattices presented in Ref.~\cite{lebren09}.
 Note however that the main 
conclusions reported in this paper are supported by a general analytical 
calculation based only on symmetry considerations, and, thus, do not depend 
on the specific details of the dynamics (\ref{eq:motion}), or if the particle
is classical or quantum. 

The quantity of interest is the directed current, formally defined as
\begin{equation}
\langle  {\bf  v}\rangle=\lim_{t\rightarrow \infty}\frac{\langle  {\bf
r}(t)-{\bf r}(0)\rangle}{t}.
\label{eq:inflimit}
\end{equation}
Such a current  is generated by the application  of an appropriate
ac force. We consider here a driving consisting of two orthogonal
bi-harmonic drives along the $x$ and $y$ directions:
\begin{subequations}\label{eq:Fxy:bih}
\begin{eqnarray}
F_x(t)&=&A_x[\cos(\omega_1 t)+\cos(2\omega_1 t+\phi_1)], \\
F_y(t)&=&A_y[\cos(\omega_2 t)+\cos(2\omega_2 t+\phi_2)],
\end{eqnarray}
\end{subequations}
with $\phi_1=\phi_2=\pi/2$.  Previous work for one-dimensional systems
has  shown  that  the  biharmonic  driving, breaking  all  the  system
symmetries,  is  able  to  produce  a  current,  whose  amplitude  and
direction can be controlled via the amplitude and the frequency of the
strength of  the driving \cite{fabio,chialvo,dykman,flayev00,machura,cubleb10,wiccub11}.  
In  the absence of coupling between transverse degrees of freedom, ac 
driving of the form of Eq.  (\ref{eq:Fxy:bih}) allows a precise control  
of transport through the 2D lattice. 

 It is important to note that,  numerically or in an experiment,
the limit (\ref{eq:inflimit}) cannot  be carried out to infinity, but
to a sufficiently large observation time $T_s$. This has important  
implications on whether  two driving frequencies $\omega_1$ and 
$\omega_2$  can be regarded as commensurate
(i.e.   $\omega_2/\omega_1$  is  a  rational  number)  or  effectively
incommensurate  (quasiperiodic  driving)  on  the time  scale  of  the
simulation.  Obviously, a periodic driving with a rational ratio 
$\omega_2/\omega_1$, specifically chosen with a period much larger than 
$T_s$,  cannot be distinguished from one with an 
irrational ratio. The periodic and quasiperiodic regimes are then 
determined  by the observation time $T_s$, as we illustrate in the 
next section. 

\section{Control of transverse coupling}
\label{sec:trans:coupling}

 In the  absence of a coupling between  the $x$ and  $y$ direction, a
driving  of the  form of  Eq. (\ref{eq:Fxy:bih})  allows a  precise
control of  transport through the  2D lattice. However, for a generic 
lattice the transverse  degrees of  freedom are
effectively coupled. This can be shown by considering the minimal case
of  a  {\it  split  biharmonic driving}  \cite{denzol08,lebren09}:
$F_x(t)=A\cos(\omega_1 t)$,  $F_y(t)=A\cos(\omega_2 t+\pi/2)$ with
$\omega_2=2\omega_1$.  For   sufficiently  large  times,   the  system
approaches an  attractor solution which is time  periodic, with period
$T=2\pi/\omega_1$.   Invariance   under  the  symmetry  transformation
$(x,y,t)\rightarrow  (-x,y,t+T/2)$  forbids  transport along  the  $x$
direction. On the  other hand, the $y$ component  of the driving force
breaks all symmetries of  the system \cite{reimann02,denzol08},
and thus  directed transport is  expected along the $y$  direction. In
our    simulations,    with     the    driving    parameters    $A=5$,
$\omega_1=\omega_2/2=\sqrt{2}$,       we       obtained       $\langle
v_x\rangle=-0.0001\pm0.0004$                and               $\langle
v_y\rangle=-0.0281\pm0.0003$,  confirming the symmetry  analysis.  The
uncertainties were estimated from  the statistics of 39000 independent
trajectories.  Note that if the system were one-dimensional, e.g along
the $y$  direction, the  single harmonic driving $F_y(t)$ would not 
induce a current, because the system would be symmetric under
the transformation $(y,t)\rightarrow (-y,t+\pi/\omega_2)$.  This analysis shows
that there  is a strong coupling between
the $x$ and  $y$ directions. The particle needs  to explore orbits the
$x$  direction  in  order to  produce  an  average  drift in  the  $y$
direction \cite{denzol08}.

As a central result of our analysis, we now show that the 
transverse coupling can be effectively suppressed by replacing the 
periodic driving considered so far by a quasiperiodic one with the 
same functional form, as obtained by choosing a driving frequency  
$\omega_2$  that  is incommensurate with  respect to $\omega_1$.
While the variation in frequency required to obtain the transition
from a periodic to a quasiperiodic driving may be tiny (few parts per 
thousand in the case studies presented in the following), the 
change in the type of driving has profound effects on the dynamics.
In fact, the transition to quasiperiodicity determines an effective
change in the dimensionality of the system. Formally, 
the compact phase space is extended \cite{neupik02} to include the variables 
$\psi_1=\omega_1 t$ and $\psi_2=\omega_2 t$.  This extension removes the 
explicit time-dependence of the problem, turning the focus from time-dependent
to stationary solutions (and thus time-periodic with period zero). Since the 
irrationality of the frequency ratio provides ergodic motion in the compact 
subspace $(\psi_1,\psi_2)$ \cite{arnold74}, it is natural to assume 
\cite{neupik02} that the dynamics in the extended phase space is ergodic. 
As a consequence, the variables $\psi_1$ and $\psi_2$ can be treated as
effectively independent variables in the symmetry analysis. 
The system, driven by the  split biharmonic force
with $\omega_2/\omega_2$ irrational, is symmetric under the transformation
$(x,y,\psi_1,\psi_2)\rightarrow(-x,-y,\psi_1+\pi,\psi_2+\pi)$,
and no directed current should appear in any direction. The
simulations  confirm  this  prediction  for  the  driving  frequencies
$\omega_1=\sqrt{2}$  and  $\omega_2=2.82$,  with  an observation  time  of
$T_s=10^5$,   resulting  in  a  zero (within the error) current with $\langle
v_x\rangle=-0.0001\pm0.0002$ and $\langle v_y\rangle=-0.0002\pm0.0003$.
 This shows that the coupling between transverse degrees of freedom
can be controlled and suppressed by using quasiperiodic ac drivings. 
Remarkably, a small variation in frequency ($\omega_1$  and $\omega_2/2$
differ in less than  0.3\%) is sufficient for the system to react as if
$\omega_2/\omega_1$  were  irrational,  displaying  a  very  different
physical    behavior   when    compared   to    the    rational   case
$\omega_1=\omega_2/2$.  The present result also represents the 
generalization to 2D of the symmetry analysis for 1D quasiperiodically 
driven systems introduced in Refs.~\cite{fladen04,gomden06,gombro07}.

\begin{figure}
\includegraphics[width=8.5cm]{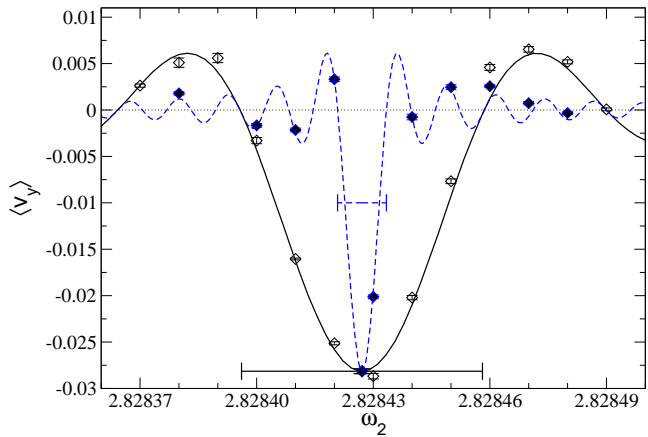}
\caption{
\label{fig:1} 
(Color online) Simulation results for the $y$  component of the current 
as a function of the frequency $\omega_2$ for a system driven by a split 
bi-harmonic driving with $\omega_1=\sqrt{2}$. The empty diamonds  
correspond to an observation time of  $T_s=10^5$, while  the  filled 
diamonds to $T_s=5\cdot10^5$.  The horizontal error bars centered at   
$\omega_2=2\sqrt{2}$  indicate $\Delta\omega=2\pi/T_s$ with   
$T_s=10^5$  (solid line) and $T_s=5\cdot10^5$ (dashed line). The  
lines are the prediction given by (\ref{eq:vypred})  with $T_s=10^5$  
(solid line)  and $T_s=5\cdot10^5$ (dashed line).}
\end{figure}

So far we only discussed the current at the exact value of the frequency 
corresponding to quasiperiodicity. For finite-time real (numerical) 
experiments, as the case considered here, it is interesting to examine the 
dependence on the current generated along the $y$ direction on the frequency 
of the control fields. Such a dependence is shown in Figure \ref{fig:1}, and it
can be precisely explained by the finite observation time $T_s$.  The 
symmetry analysis discussed earlier, which  assumes  an  infinite  $T_s$,  
predicts  that  only  the  value $\omega_2=2\sqrt{2}$  of those  shown in  
Fig.~\ref{fig:1}  produces a current different from zero. Correspondingly, 
the Fourier cosine transform of the  single harmonic $F(t)=\cos(\omega_0  t)$ 
is proportional  to a Dirac delta centered at $\omega_0$,
\begin{equation}
\int_0^\infty dt\,\cos(\omega_0 t)\cos(\omega t)=\pi\delta(\omega-\omega_0).
\end{equation}
However, when the finite observation time $T_s$ is taken into account, the
Fourier transform has to be replaced by
\begin{equation}
\int_0^{T_s}              dt\,\cos(\omega_0              t)\cos(\omega
t)=\pi\delta_{1/T_s}(\omega+\omega_0)+\pi\delta_{1/T_s}(\omega-\omega_0),
\label{eq:deltats}
\end{equation}
where $\delta_{\epsilon}(x)=\mathrm{sinc}(x/\epsilon)/(\pi\epsilon)$ is a well
known  representation of  the delta  function with  the  sinc function
$\mathrm{sinc}(x)=\sin(x)/x$.  The  first  delta function of  the  right-hand side  of
(\ref{eq:deltats}) is irrelevant because the frequency $\omega$ in the
Fourier cosine transform is  only defined for $\omega\ge0$. Therefore,
we  would  expect  that  the  system response  also  shows  a  similar
frequency broadening  in the neighborhood  of $\omega_2=2\omega_1$ due
to the finite duration $T_s$,
\begin{equation}
\langle v_y\rangle\approx v_0\cdot\mathrm{sinc}[(\omega_2-2\omega_1)T_s],
\label{eq:vypred}
\end{equation}
where   $v_0$   is   the    value   of   $\langle   v_y\rangle$   when
$\omega_2=2\omega_1$   (note  that   $\mathrm{sinc}(0)=1$).    The  lines   in
Fig.~\ref{fig:1}  show  that  the  shape  is well  described  by  this
approximation. The width $\Delta \omega$ of the resonance around the 
value which defines quasiperiodicity is simply the frequency  
resolution introduced by  the finite duration  $T_s$ of
the measurement: $\Delta\omega=2\pi/T_s$.  For a real experiment, 
such a width controls the frequency window within which the driving
could be regarded approximately as periodic, i.e. it defines the 
frequency jump required to move from the periodic driving regime to the
quasiperiodic one. 

\section{Control of transport in 2D with quasiperiodic driving}

\begin{figure}
\includegraphics[width=8.5cm]{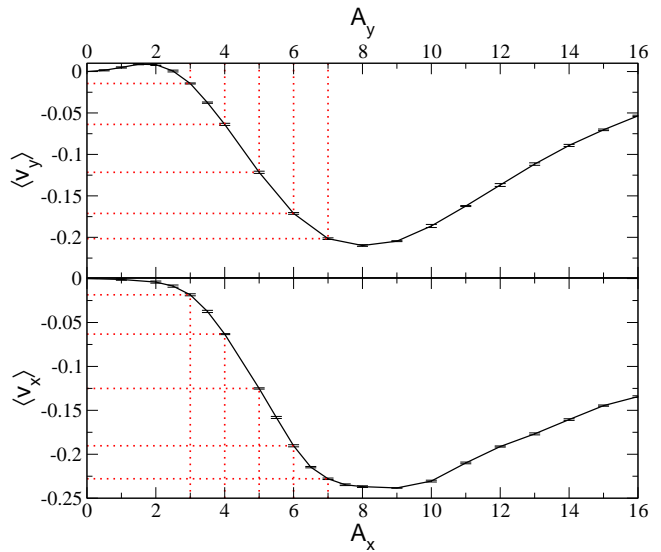}
\caption{
\label{fig:2} 
(Color online) Simulation results for the 1D current, as obtained by applying a 
biharmonic driving along one direction only, as a function  of the 
amplitude of the  driving for the relevant direction. Using the 
notations as from Eq. (\ref{eq:Fxy:bih}), the parameters of 
the calculations are as follows. 
Top  panel: $\langle  v_y\rangle$ vs.  $A_y$ for  $A_x=0$; bottom  panel: 
$\langle v_x\rangle$  vs. $A_x$ for $A_y=0$. In all cases
$\omega_1=\omega_2=\sqrt{2}$.  The  solid lines  are  a  guide to  the
eye. The  dotted lines (red) mark a  set of reference  values of  $A_x$ and
$A_y$.}
\end{figure}

We now consider the problem of the control of transport in 2D with 
ac drivings. We analyze the simplest case of drivings breaking all the
relevant symmetries, the double biharmonic driving, Eq. (\ref{eq:Fxy:bih}).

Previous work \cite{lebren09} demonstrated that it is possible to produce 
directed motion along an arbitrary direction of the 2D substrate by 
using ac driving forces. However, the mechanism shown in that work lacks the
essential feature of a control protocol: predictability. Indeed,
because of the coupling between transverse degrees of freedom, and the 
nonlinearity of the mechanism of rectification along each direction,
it is impossible, given the parameters of the driving, to predict in a 
straightforward way the direction along which directed motions will be 
produced. Only a complete calculation, which also requires the exact 
knowledge of the geometry of the 2D structure, can reveal the direction 
of the current which is in general different from the direction corresponding 
to the vector sum of the forces oscillating in the two directions. 

As it will be shown here, the use of quasiperiodic ac fields leads instead
to a simple control protocol, which produces a current closer to a direction 
corresponding to the vector sum of the forces oscillating in the two 
directions, independently of the lattice geometry. 

As a starting point, we consider the 1D current, as obtained by applying a
biharmonic driving along one direction only. Numerical results for this case  
are reported in Fig.~\ref{fig:2}. The observation time was fixed to $T_s=10^5$.
Two general remarks are in order. First, the sign of the current (negative
for the considered parameters) is not important as it can be controlled
by inverting the values of $A_x$ and/or $A_y$  or   changing  the  
values of $\phi_1$ and/or $\phi_2$ to $\phi_1\rightarrow \phi_1 +\pi$ 
and  $\phi_2\rightarrow \phi_2 +\pi$.  In either  case, the sign of the 
current component $\langle v_x\rangle$ and/or $\langle v_y\rangle$ would be 
reversed. Second, it can be seen  that for the relatively  small values of
the driving amplitudes (about $A_x,A_y\le 2$ in Fig.~\ref{fig:2}), the
current  remains very  small. A  functional
expansion  on the driving  amplitude confirms that no current is generated  
at the  first \cite{reimann02}  (linear response  theory) and
second order on the driving amplitude \cite{quicue10}.
Furthermore, Fig.~\ref{fig:2} shows that the current in each direction
presents a non-monotonous behavior with the driving amplitudes (showing
two minima at about $A_x, A_y \sim 8$). This is also expected, because
for very large  driving amplitudes the potential can  be neglected and
thus, the potential's nonlinearity, which determines the current 
generation, diminishes,  eventually leading  to the  disappearance  
of the current for large enough driving.
Since we  are interested in  controlling the directed  current through
the driving amplitudes,  it will suffice to restrict  ourselves to the
range of  parameter values  defined by $3\le  A_x,A_y \le 7$.

\begin{figure}
\includegraphics[width=8.5cm]{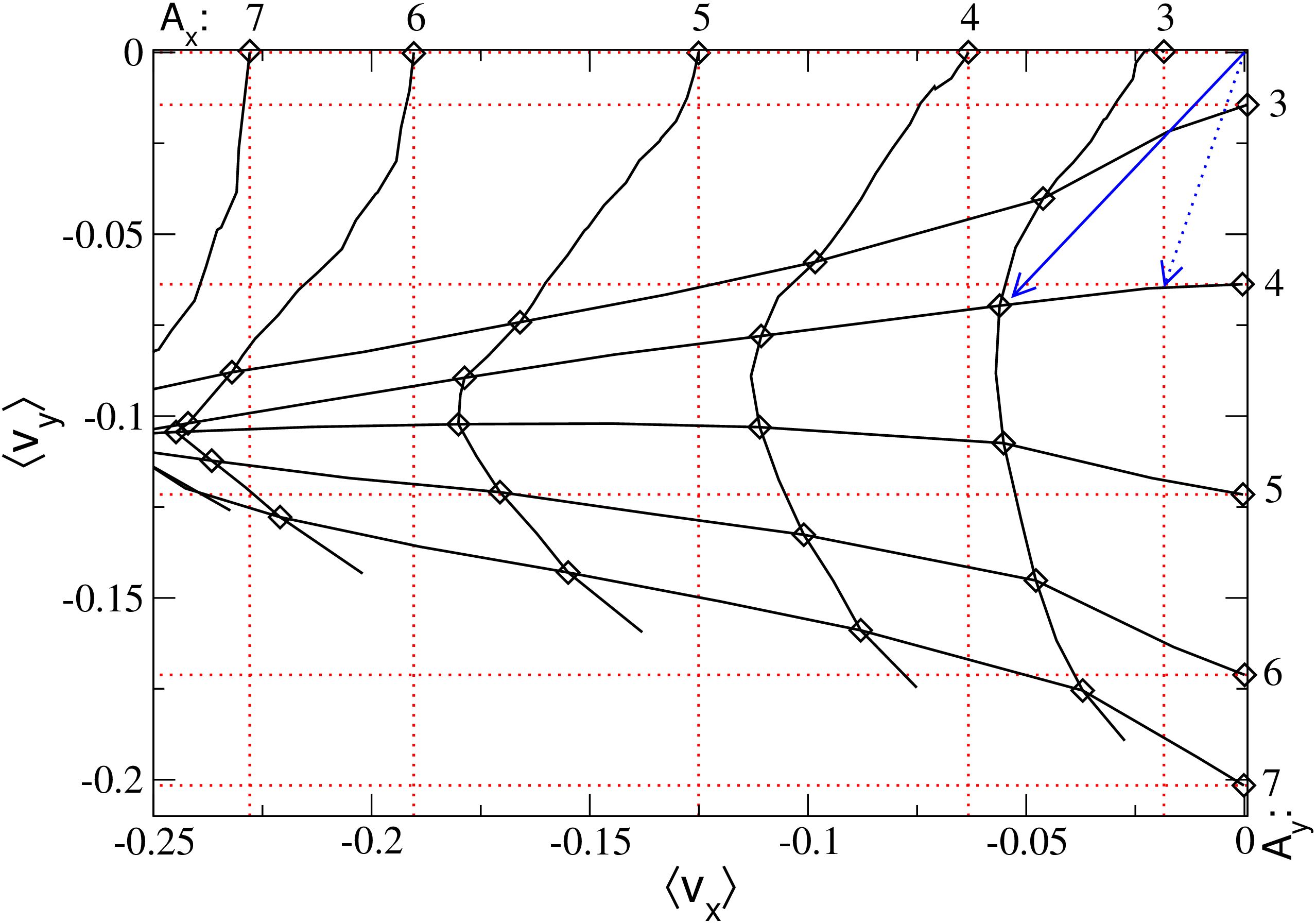}
\caption{
\label{fig:3} 
(Color online) Simulation  results for the rectangular potential of 
Eq.~(\ref{eq:rect_pot}) and the bidimensional driving of Eq.~(\ref{eq:Fxy:bih})
with $\omega_1=\omega_2=\sqrt{2}$. The numbers  on the top of the plot
mark  the values  of $A_x$  for each  {\em vertical}  (constant $A_x$) 
solid  line. The numbers on  the right axis  of the plot  mark the values 
of  $A_y$ for each  {\em horizontal}  (constant $A_y$)  solid  line. 
The  dotted (red) lines  are a  guide to indicate the  current values marked in 
Fig.~\ref{fig:2},  and thus the values that would be obtained if we could 
neglect the coupling between both directions. The dotted and solid (blue) 
arrows indicate the direction of the current for $(A_x,A_y)=(3,4)$ in the 
ideal and real cases, respectively. }
\end{figure}

\begin{figure}
\vspace{0.5em} \includegraphics[width=8.5cm]{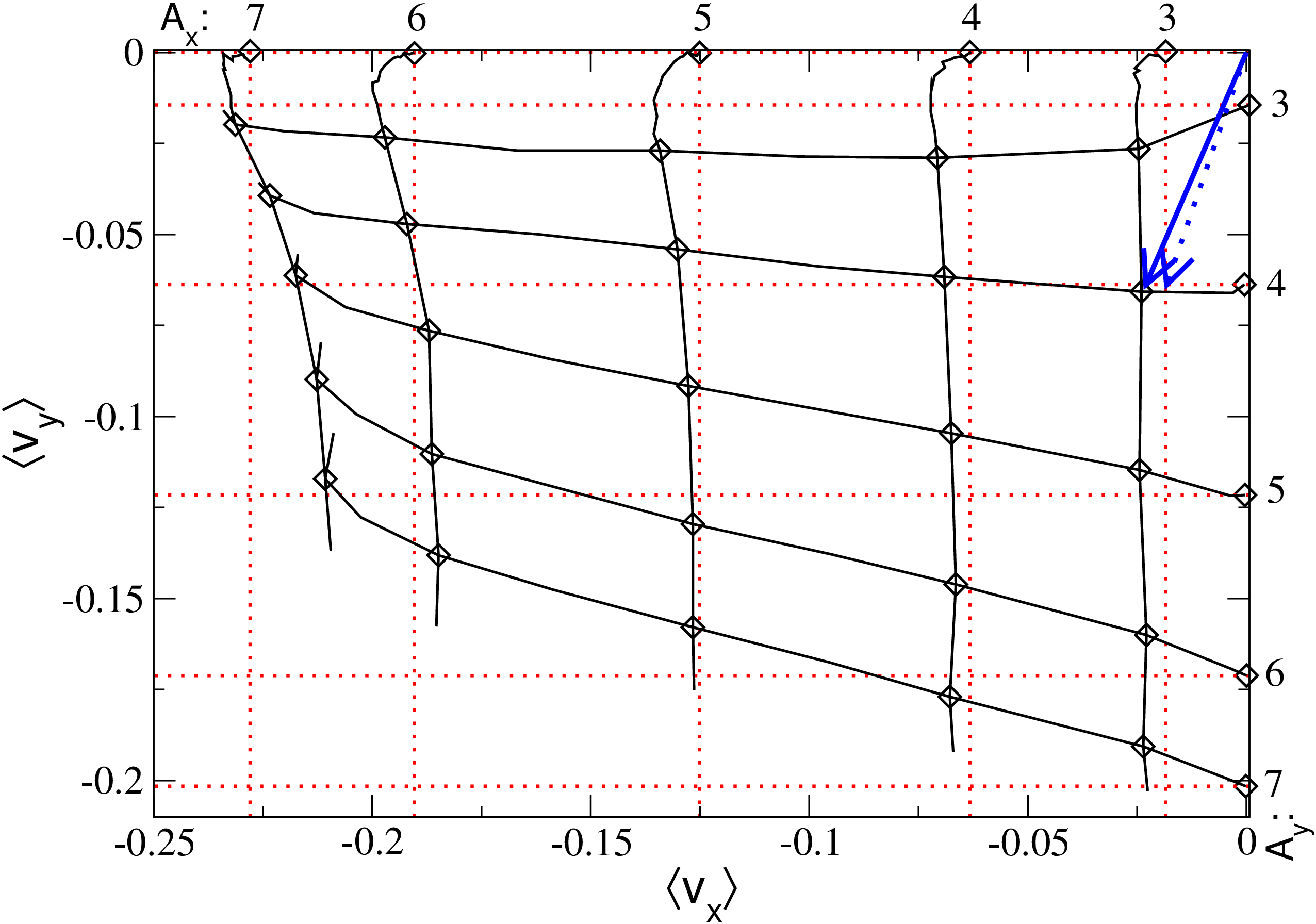}
\caption{
\label{fig:4} 
(Color online) Same as  in   Fig.~\ref{fig:3}  but   for   $\omega_1=\sqrt{2}$  
and $\omega_2=1.41$,    showing    a    considerably    reduced    lattice
deformation.  Note that  the  difference here  between $\omega_1$  and
$\omega_2$ is less than 0.3\%.}
\end{figure}

If we intend to produce a  current in a direction other than along the
axes, we need to simultaneously apply drivings in both $x$ and $y$ directions.
Fig.~\ref{fig:3} shows what  happens when  this is
done.  The ideal  situation  for  direction control  would  be that  a
superposition  principle  would apply,  so  that  a specific  required
current  direction could  be  obtained by  applying the  corresponding
driving   amplitudes  in   each   perpendicular  direction.   However,
Fig.~\ref{fig:3} shows a very large deviation from this behavior, with the
directed  current values  (solid lines  and diamonds  at  the crossing
between  the  lines)  going  far  away from  the  ideal  case  (dotted
lines).   Looking   for   example   at   the   current   produced   at
$(A_x,A_y)=(3,4)$, one would expect, after observing the corresponding
values  at Fig.~\ref{fig:2} (which are indicated in Fig.~\ref{fig:3} 
with dotted lines),  that a current is formed along the direction 
indicated in  Fig.~\ref{fig:3} by the dotted arrow. However, the current 
ends up having the direction given by the solid arrow, which forms a much 
larger angle with the $y$ axis than expected. In addition, further increasing
$A_y$  additionally produces  an  unexpected non-monotonous behavior  in 
$\langle v_x\rangle $, which makes control of the current direction rather difficult.

This phenomenon is due to the  strong coupling between the $x$ and $y$
components       at      the       same       driving      frequencies
$\omega_1=\omega_2$. Remarkably, we can significantly suppress this coupling
by using two incommensurate frequencies,  as shown  in
Fig.~\ref{fig:4}. Note that the difference in $\omega_2$ with the case
of periodic driving shown in Fig.~\ref{fig:2} is just less than 0.3\%,
which implies that the curve shown  in the top panel of this figure is
practically indistinguishable  from the  one obtained with  the latter
frequency $\omega_2=1.41$.  Fig.~\ref{fig:4} shows that the deviation from
an ideal behavior of uncoupled $x$ and $y$ dynamics is significantly
reduced, in particular for weak driving [$A_{x,y} \le 5$]. The deviation 
from  such an ideal behavior is still pronounced at larger
driving fields, with driving amplitude values close to the minima shown in Fig.~\ref{fig:2}. 

\begin{figure}
\includegraphics[width=8.5cm]{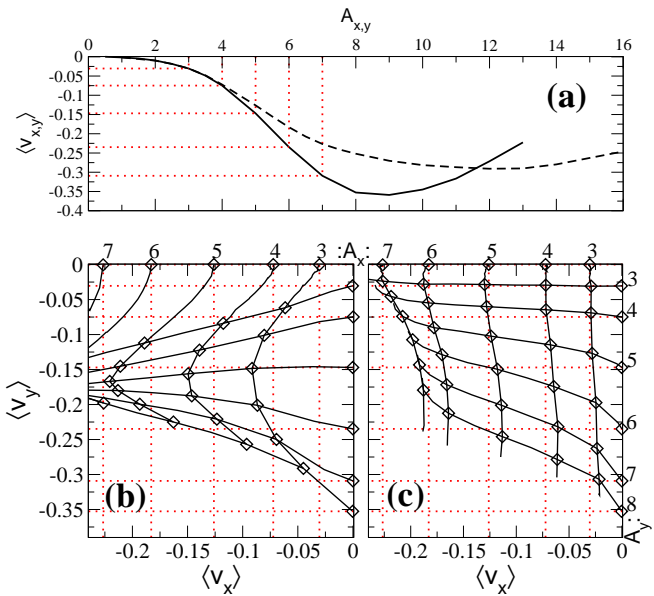}
\caption{
\label{fig:5} 
(Color online) Simulation  results for the hexagonal potential 
(\ref{eq:hex_pot}) and the bidimensional  driving (\ref{eq:Fxy:bih}) with 
$\omega_1=\sqrt{2}$ and: (b) $\omega_2=\sqrt{2}$ (periodic driving) and (c) 
$\omega_2=1.41$ (quasiperiodic driving). 
(a) shows the 1D current, when the biharmonic driving is applied in one 
direction only ($x$: solid line and $y$: dashed line), as a function of the 
amplitude the driving in the relevant direction.
}
\end{figure}

A similar behavior is observed for a system with the following potential
\begin{equation}\label{eq:hex_pot}
U({\bf r})=U_0\left[\cos(k x)+2\cos(kx/2)\cos(\sqrt{3}y/2)\right],
\end{equation}
which produces an {\em hexagonal lattice} in the XY plane, being in addition 
spatially symmetric in both perpendicular directions. Fig.~\ref{fig:5} 
shows that the decoupling produced by the quasiperiodic driving is almost 
perfect for small driving amplitudes, allowing a precise control of the 
current direction. 

\begin{figure}
\includegraphics[width=8.5cm]{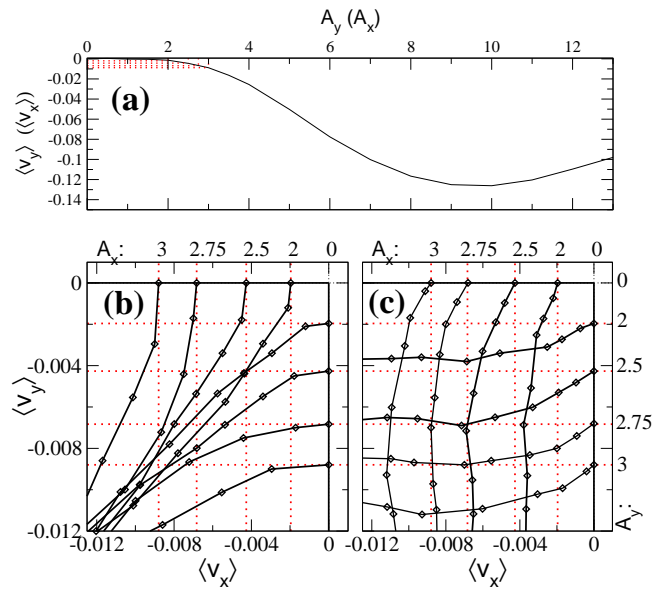}
\caption{
\label{fig:6} 
(Color online) Simulation  results for the square potential 
of Eq.~(\ref{eq:square_pot}) and the bidimensional  driving of 
Eq.~(\ref{eq:Fxy:bih}) with $\omega_1=\sqrt{2}$ and: (b) $\omega_2=\sqrt{2}$ 
(periodic driving) and (c) $\omega_2=1.41$ (quasiperiodic driving). 
(a) shows the 1D current when the biharmonic driving is applied in 
one direction only as a function of the amplitude the driving in the 
relevant direction. In (b) and (c) the simulation data is represented 
by diamonds, with the solid lines being a guide to the eye connecting 
the data points. Each data point is estimated to have an error bar of 
about $\Delta v=0.001$ in each perpendicular direction (not drawn for clarity). }
\end{figure}

We have also studied a {\em square lattice}. Figure~\ref{fig:6} shows the simulation results for the potential 
\begin{equation}\label{eq:square_pot}
U({\bf r})=U_0\cos(k x)\cos(ky).
\end{equation}
Due to the explicit symmetry in the potential between the $x$ and $y$
directions, the directed current displays the strongest couplings when
the biharmonic driving  (\ref{eq:Fxy:bih}) is applied in both
directions. The coupling is so strong that no significant improvement
is found even with the quasiperiodic driving for moderate values of
the driving amplitudes. Only at very small driving amplitudes -- the
values indicated in Fig.~\ref{fig:6}a with dotted lines -- the
quasiperiodic driving is able to diminish the couplings so that a
reasonable control of the current direction is possible. Note that the
current values shown in Fig.~\ref{fig:6}a and \ref{fig:6}b are very
small, and the simulation error bars are thus of considerable
size. Still, it can be observed that the quasiperiodic driving is able
to reduce significantly the large lattice distortion produced by the
couplings.

In fact, we prove in the Appendix that this is a general result
applicable to any spatially periodic system that is also spatially
symmetric in both the $x$ and $y$ directions. A functional
expansion in the driving amplitudes shows that the directed
current of a system driven by the forces (\ref{eq:Fxy:bih}) with
$\omega_2/\omega_1$ irrational is, in the first orders in the
driving amplitudes $A_x$ and $A_y$,
\begin{subequations}\label{eq:vxvy:expansion}
\begin{eqnarray}
\langle v_x\rangle=A_x^3 B_{x0}\cos(\phi_1-\phi_{x0})+ \mathcal{O}(5), 
\\
\langle v_y\rangle=A_y^3 B_{y0}\cos(\phi_2-\phi_{y0})+ \mathcal{O}(5),
\end{eqnarray}
\end{subequations}
where  $B_{x0}$ ($B_{y0}$) and $\phi_{x0}$ ($\phi_{y0}$) are
independent of the driving parameters $A_x$, $A_y$, $\phi_1$,
$\phi_2$, and $\omega_2$ ($\omega_1$). Explicit expressions for the
fifth order terms are given in the appendix. Therefore, in the lowest
order on the driving amplitudes ($A_x^3$ for $\langle v_x\rangle$ and
$A_y^3$ for $\langle v_y\rangle$), the current contains no coupling
between the $x$ and $y$ directions when the quasiperiodic driving is
applied.  In contrast, with the periodic driving $\omega_1=\omega_2$,
the current contains additional third order terms such as $A_xA_y^2$
in $\langle v_x\rangle$ and $A_yA_x^2$ in $\langle v_y\rangle$ (see
the appendix), which makes the control of the current direction rather
difficult for any values of the driving amplitudes. These
considerations are not restricted to the specific equation of motion
(\ref{eq:motion}), since the calculations rely only on 
general symmetry considerations.

 The observed partial loss of control at large 
driving amplitudes can also be explained  within the framework of 
Dynamical Systems theory \cite{dynsys1,dynsys2}.   
The robustness of a quasiperiodic state can be understood 
by considering the two phases $\psi_1$, $\psi_2$ as coupled
\cite{dynsys1}.
We refer to the exactly solvable model 
\begin{equation}
\dot{\psi}_1=\omega_1+f_1(\psi_1,\psi_2), \quad
\dot{\psi}_2=\omega_2+f_2(\psi_1,\psi_2),
\end{equation} 
with $\omega_1$, $\omega_2$ incommensurate frequencies, and 
 $f_1$ and $f_2$ arbitrary coupling functions that are
$2\pi$-periodic in each argument. This  model is useful
to highlight the loss of quasiperiodicity in the dynamics at large 
driving amplitudes which, in our system, leads to loss of control. 
The key observation is the dependence of the commensurability of 
the observed frequencies
$\Omega_1=\langle \dot{\psi}_1\rangle$ and $\Omega_1=\langle
\dot{\psi}_2\rangle$ on the coupling functions.  It is known 
that for small coupling $f_{1,2}$, the measure of all parameter values 
for which periodic regimes (i.e. $\Omega_1$ and $\Omega_2$ commensurate) 
are observed is small, while the measure of the corresponding quasiperiodic 
states is large. For large $f_{1,2}$, the measure of the periodic regimes grows, 
while that of quasiperiodic regimes decreases. These features are 
in agreement with the observed behavior in the 2D driven systems 
studied here.

\section{Conclusions}
In conclusion, in this work we consider the problem of the control
of  transport in higher dimensional periodic structures by applied ac
fields.  In a generic lattice, transverse degrees of freedom are
coupled,  and this makes the control of motion difficult to
implement. We show,  both with a numerical and a rather
general analytical analysis, that the use of quasiperiodic driving
significantly suppresses the coupling between  transverse degrees of
freedom. Remarkably, this requires tiny variations  of the frequency
of the control field, of the order of few parts per  thousand for the
case studies presented in this work. The specific minimum
variation required for quasiperiodic behavior in a real experiment or
simulation is shown to depend on the observation time, as expected. 

The dynamical decoupling of degrees of freedom allows a precise
control of the  transport, and does not require a detailed knowledge
of the crystal geometry. Our results are of  relevance for the control
of transport in higher dimensional systems in which direct control, or
knowledge, of the substrate  geometry is lacking, as  usually
encountered in solid state systems \cite{vortex}.

\begin{acknowledgments}
This research was funded by  the  Leverhulme Trust, and the 
Ministerio  de Ciencia e Innovaci\'on of Spain FIS2008-02873 (DC).
\end{acknowledgments}

\appendix

\section{Functional expansion in the driving amplitudes}
We follow here the powerful method presented in Ref.~\cite{quicue10}
for a one-dimensional spatially periodic and symmetric system subject
to the driving force
\begin{equation}
F(t)=A[\cos(p\omega t+\varphi_1)+\cos(q\omega t+\varphi_2)],
\end{equation}
where $p$ and $q$ are positive integers. The current $\langle
v\rangle=v[F]$ has a functional dependence on $F(t)$, and thus, it can
be Taylor expanded as
\begin{eqnarray}\label{eq:ap:taylor1D}
v[F]&=&\sum_{n\ge0}v_n[F],\nonumber\\
v_n[F]&=&\{c_n(t_1,\ldots,t_n)F(t_1)\cdots F(t_n)\},
\end{eqnarray}
where
\begin{equation}
 \{f(t_1,\ldots,t_n)\}\equiv\frac{1}{T^n}\int_0^Tdt_1\cdots\int_0^Tdt_n\,f(t_1,\ldots,t_n),
\end{equation}
$T$ is the period of the driving, and $c_n$ functions that can be
chosen totally symmetric under any exchange of their arguments. It is
shown in \cite{quicue10} that, when the system symmetries are taken
into account, all terms in (\ref{eq:ap:taylor1D}) with  $n<p+q$
vanish, giving the lowest order possible contribution at
$n=p+q$. Therefore, in the {\em quasiperiodic limit}, defined as
$p,q\rightarrow \infty$ and $\omega\rightarrow 0$, so that
$q\omega=\omega_1, p\omega=\omega_2$, with $\omega_1$ and $\omega_2$
two incommensurate (finite) frequencies, all terms in
(\ref{eq:ap:taylor1D}) vanish, producing the expected suppression of
current.

We can apply this method to a 2D system by using the expansion
(\ref{eq:ap:taylor1D}) for any component of the current, and then
further Taylor expanding $c_n$ on the other component of the driving
force. We then obtain for the $x$ component $\langle
v_x\rangle=v_x[{\bf F}]$,
\begin{subequations}\label{eq:ap:taylor2D}
\begin{eqnarray}
v_x[{\bf
F}]&=&\sum_{n_x\ge1}v_{n_x,0}^{(x)}[F_x,0]+\sum_{n_xn_y\ge1}v_{n_x,n_y}^{(x)}[{\bf
F}],\label{eq:ap:taylor2D:sum}\\ v_{n_x,n_y}^{(x)}[{\bf
F}]&=&\{c_{n_x,n_y}(t_1,\ldots,t_{n_x};t'_1,\ldots,t'_{n_y})\nonumber\\
& &\times F_x(t_1)\cdots F_x(t_{n_x})F_y(t'_1)\cdots
F_y(t'_{n_y})\},\nonumber\\
\end{eqnarray}
\end{subequations}
where we have already used the fact that $v_x[0,F_y]=0$ because of the
system symmetries (see Eq.~(\ref{eq:vx:symm:comp})), and thus excluded
the possibility $n_x=0$ from (\ref{eq:ap:taylor2D:sum}). The first sum
in the right-hand side of (\ref{eq:ap:taylor2D:sum}) contains the terms
which are independent of the transverse driving component $F_y(t)$,
while the second sum accounts for the transverse couplings.

 Before continuing, let us state  explicitly the basic symmetries 
 that we are going to use in the calculation. First, the potential $U(x,y)$ 
 must be spatially symmetric in both directions, i.e. for each $y$ ($x$), 
 there must exist a $x_0$ ($y_0$) such as
\begin{subequations}
\begin{eqnarray}
U(x_0+x,y)=U(x_0-x,y) \mbox{ for all }x,\\
U(x,y_0+y)=U(x,y_0-y) \mbox{ for all }y.
\end{eqnarray}
\end{subequations}
In this situation, the current can only appear by the application of a
symmetry-breaking driving force, which thus controls the sign of the current
\begin{subequations}
\begin{eqnarray}
v_x[-F_x,-F_y]=-v_x[F_x,F_y],\label{eq:vx:symm}\\
v_y[-F_x,-F_y]=-v_y[F_x,F_y],
\end{eqnarray}
\end{subequations}
and for each component,
\begin{subequations}
\begin{eqnarray}
v_x[-F_x,F_y]=-v_x[F_x,F_y],\label{eq:vx:symm:comp}\\
v_y[F_x,-F_y]=-v_y[F_x,F_y].
\end{eqnarray}
\end{subequations}
To satisfy the condition (\ref{eq:vx:symm}), the functions
$c_{n_x,n_y}$ in (\ref{eq:ap:taylor2D}) have to be identically zero
for even values of $n=n_x+n_y$. Similarly, (\ref{eq:vx:symm:comp})
implies no contribution in (\ref{eq:ap:taylor2D}) from terms with even
values of $n_x$.  In addition, in dissipative systems,  as the one considered here,  the 
current usually does not depend on the specific choice of time origin,
\begin{subequations}\label{eq:symm:t0}
\begin{eqnarray}
v_x[{\bf F}(t+t_0)]&=&v_x[{\bf F}(t)],\\
v_y[{\bf F}(t+t_0)]&=&v_y[{\bf F}(t)],
\end{eqnarray}
\end{subequations}
for any $t_0$. In non-dissipative systems displaying a strong dependence on the initial conditions, as in Hamiltonian ratchets \cite{reimann02}, the condition (\ref{eq:symm:t0}) can generally be satisfied either by averaging over the initial time \cite{flayev00}, or by adiabatically switching on the driving ${\bf F}(t)$. The implications of (\ref{eq:symm:t0}) depend on the
explicit form of the driving force. Instead of (\ref{eq:Fxy:bih}), let
us consider the following -- slightly more general -- biharmonic driving
\begin{subequations}\label{eq:apen:Fxy:bih}
\begin{eqnarray}
F_x(t)&=&A_x[\cos(\omega_1 t+\hat{\phi}_1)+\cos(2\omega_1 t+\phi_1)], \\
F_y(t)&=&A_y[\cos(\omega_2 t+\hat{\phi}_2)+\cos(2\omega_2 t+\phi_2)],
\end{eqnarray}
\end{subequations}
where $\hat{\phi}_1$ and $\hat{\phi}_2$ are new driving phase
constants. The conditions (\ref{eq:symm:t0}) imply that the current
must be invariant under the following transformation
\begin{eqnarray}\label{eq:apen:transf_t0}
\hat{\phi}_1\rightarrow\hat{\phi}_1+\omega_1t_0, \quad \phi_1\rightarrow\phi_1+2\omega_1t_0, \nonumber\\
\hat{\phi}_2\rightarrow\hat{\phi}_2+\omega_2t_0, \quad \phi_2\rightarrow\phi_2+2\omega_2t_0,
\end{eqnarray}
for any arbitrary $t_0$.

Expanding the cosines in (\ref{eq:apen:Fxy:bih}) in complex exponentials yields
\begin{eqnarray}\label{eq:apen:vnxny:expan}
v_{n_x,n_y}^{(x)}[{\bf F}]=\sum_{{\bf n}\ge0}^\otimes A_x^{n_x} A_y^{n_y}C({\bf n})e^{i\Theta({\bf n},{\boldsymbol \phi}) },
\end{eqnarray}
where ${\bf n}=(n_1,n_2,n_3,n_4,n_1',n_2',n_3',n_4')$, the symbol
$\otimes$ denotes a restriction in the sum to the values of the
tuple ${\bf n}$ such that
\begin{eqnarray}\label{eq:apen_sum:restr}
n_1+n_2+n_3+n_4=n_x,\nonumber\\
n'_1+n'_2+n'_3+n'_4=n_y,
\end{eqnarray}
${\bf n}\ge0$ denotes a component-wise inequality, 
\begin{equation}
{\boldsymbol \phi}=(\hat{\phi}_1,\phi_1,\hat{\phi}_2,\phi_2),
\end{equation}
 and
\begin{eqnarray}
\Theta({\bf n},{\boldsymbol \phi})&=&[(n_1-n_2)\hat{\phi}_1+(n_3-n_4)\phi_1\nonumber\\
& &+(n_1'-n_2')\hat{\phi}_2+(n_3'-n_4')\phi_2].
\end{eqnarray}
$C$ is a complex function of ${\bf n}$, $\omega_1$ and $\omega_2$ that
can be traced back to time integrals of $c_{n_x,n_y}$ multiplied by
the factors $e^{\pm i\omega_1t_k}$ and  $e^{\pm
i\omega_2t_k'}$. Further, it satisfies $C(\hat{\bf n})=C({\bf
n})^\ast$, where $\ast$ denotes complex conjugate, and
\begin{equation}
\hat{\bf n}=(n_2,n_1,n_4,n_3,n'_2,n'_1,n'_4,n'_3).
\end{equation}
Thus, for every term in (\ref{eq:apen:vnxny:expan}) with tuple 
${\bf n}$, there is another term given by $\hat{\bf n}$ which is 
just the complex conjugate of the former, guaranteeing that 
$v_{n_x,n_y}^{(x)}[{\bf F}]$ is real. 

From Eq.~(\ref{eq:apen:vnxny:expan}), it is clear that the order of 
$v_{n_x,n_y}^{(x)}[{\bf F}]$ is given by the factor $A_x^{n_x} A_y^{n_y}$, 
and thus by  $n=n_x+n_y$.

Notice that the transformation (\ref{eq:apen:transf_t0}) only 
affects $\Theta$ in (\ref{eq:apen:vnxny:expan}). More specifically, 
it implies
\begin{equation}\label{eq:apen:transf_t0:2}
(n_1-n_2)+2(n_3-n_4)+\frac{\omega_2}{\omega_1}\left[(n'_1-n'_2)+2(n'_3-n'_4)\right]=0.
\end{equation}
Since $\omega_2/\omega_1$ is an irrational number, Eq.~(\ref{eq:apen:transf_t0:2}) is 
only satisfied when
\begin{eqnarray}\label{eq:apen:transf_t0:3}
(n_1-n_2)+2(n_3-n_4)=0,\nonumber\\
(n'_1-n'_2)+2(n'_3-n'_4)=0.
\end{eqnarray}
The restrictions (\ref{eq:apen:transf_t0:3}), together with  (\ref{eq:apen_sum:restr}) 
and the above mentioned conditions in $n_x$ and $n$ given by (\ref{eq:vx:symm}) and 
(\ref{eq:vx:symm:comp}), determine the possible terms in the expansion (\ref{eq:ap:taylor2D:sum}). 

The lowest level in the expansion satisfying the above conditions is given by 
$v_{3,0}^{(x)}[F_x,0]$, which is obviously independent of $F_y(t)$, having a 
contribution coming from the tuple ${\bf n}=(2,0,0,1,0,0,0,0)$ (and its corresponding complex 
conjugate $\hat{\bf n}$), and thus
\begin{equation}
v_{3,0}^{(x)}[F_x,0]=A_x^3 B_{x0}\cos(\phi_1-2\hat{\phi}_1-\phi_{x0}),
\end{equation}
where $B_{x0}$ and $\phi_{x0}$ depend on the driving parameters only
through $\omega_1$. This is the only third order term satisfying
(\ref{eq:apen:transf_t0:3}). All fourth order terms are forbidden due
to the symmetry (\ref{eq:vx:symm}). In the fifth order, there is one
term containing no transverse coupling, $v_{5,0}^{(x)}$, coming from
the tuples ${\bf n}=(2,0,1,2,0,0,0,0)$ and $(3,1,0,1,0,0,0,0)$. Then,
\begin{eqnarray}
v_{5,0}^{(x)}[F_x,0]&=&A_x^5[B_{x1}\cos(\phi_1-2\hat{\phi_1}-\phi_{x1})\nonumber\\
& &+B_{x2}\cos(\phi_1-2\hat{\phi_1}-\phi_{x2})],
\end{eqnarray}
where $B_{xj}$ and $\phi_{xj}$, with $j=1,2$, depend on $\omega_1$
only.  In this order, the only surviving coupling term is given by
$v_{3,2}^{(x)}$, which has contributions from the tuples ${\bf
n}=(2,0,0,1,1,1,0,0)$ and $(2,0,0,1,0,0,1,1)$, yielding
\begin{eqnarray}
v_{3,2}^{(x)}[F_x,F_y]&=&A_x^3A_y^2[B'_{x1}\cos(\phi_1-2\hat{\phi_1}-\phi'_{x1})\nonumber\\
& &+B'_{x2}\cos(\phi_1-2\hat{\phi_1}-\phi'_{x2})],
\end{eqnarray}
where now $B'_{xj}$ and $\phi'_{xj}$ depend on $\omega_1$ and $\omega_2$.

Finally, note that when the ratio $\omega_2/\omega_1$ is rational (the case 
of periodic driving) there are additional terms that satisfy 
(\ref{eq:apen:transf_t0:2}). More specifically, for $\omega_1=\omega_2$ the 
coupling term $v_{1,2}^{(x)}$ gives a non-vanishing contribution from the 
tuples ${\bf n}=(1,0,0,0,1,0,0,1)$ and $(0,0,0,1,2,0,0,0)$,
\begin{eqnarray}
v_{1,2}^{(x)}[F_x,F_y]&=&A_xA_y^2[B_{x1}^p\cos(\phi_2-\hat{\phi_1}-\hat{\phi_2}-\phi_{x1}^p)\nonumber\\
& &+B_{x2}^p\cos(\phi_1-2\hat{\phi_2}-\phi_{x2}^p)].
\end{eqnarray}

 

\end{document}